\documentclass[conference]{IEEEtran}

\usepackage{amsmath}
\usepackage{amsfonts}
\usepackage{amssymb}
\usepackage{amsthm}
\usepackage{epsfig}
\usepackage{latexsym}
\usepackage{comment}
\usepackage{graphicx}
\usepackage{float}
\usepackage[leftcaption]{sidecap}
\usepackage{xspace}
\usepackage[section]{algorithm}
\usepackage[noend]{algorithmic}

\newtheorem{theorem}{Theorem}
\newtheorem{lemma}[theorem]{Lemma}

\newtheorem{cor}[theorem]{Corollary}

\theoremstyle{definition}

\theoremstyle{remark}

\newcommand{\cm}[1]{}

\newcommand{\cL}{{\cal L}}
\newcommand{\cC}{{\cal C}}
\newcommand{\cR}{{\cal R}}
\newcommand{\cN}{{\cal N}}

\begin{document}

\title{\huge{Utility Optimization in Heterogeneous Networks via CSMA-Based Algorithms}
}
\author{\IEEEauthorblockN{Matthew Andrews}
\IEEEauthorblockA{
Alcatel-Lucent Bell Labs, Murray Hill, NJ \\
andrews@research.bell-labs.com}
\and
\IEEEauthorblockN{Lisa Zhang}
\IEEEauthorblockA{
Alcatel-Lucent Bell Labs, Murray Hill, NJ \\
ylz@research.bell-labs.com}
\vspace{-1in}
}
\date{}
\maketitle

\begin{abstract}
We study algorithms for carrier and rate allocation in cellular
systems with distributed components such as a heterogeneous LTE system
with macrocells and femtocells.  Existing work on LTE systems often
involves centralized techniques or requires significant signaling, and 
is therefore not always applicable in the presence of femtocells.
More distributed CSMA-based algorithms (carrier-sense multiple access)
were developed in the context of 802.11 systems and have been proven
to be utility optimal. However, the proof typically assumes a single
transmission rate on each carrier. Further, it relies on the CSMA
collision detection mechanisms to know whether a transmission is
feasible.

In this paper we present a framework for LTE scheduling that is based
on CSMA techniques. In particular we first prove that CSMA-based
algorithms can be generalized to handle multiple transmission rates in
a multi-carrier setting while maintaining utility optimality.  We then
show how such an algorithm can be implemented in a heterogeneous LTE
system where the existing Channel Quality Indication (CQI) mechanism
is used to decide transmission feasibility.
\end{abstract}

\section{Introduction}

Interference mitigation is a fundamental problem in wireless networks.
The exact method for handling interference depends on the nature of
the network, e.g. whether it is a centrally controlled cellular
network or a more unstructured ad-hoc network.  For cellular networks
interference can be mitigated via techniques such as power-control,
frequency reuse, and fine-grained rate control based on
channel-quality measurements together with some aspect of central
planning.  On the other hand, for ad-hoc networks, especially those
running the 802.11 protocol, interference is typically mitigated by a
distributed collision-based random access scheme, perhaps coupled with
a fairly coarse-grained rate-adaptation procedure.

In this paper we are concerned with interference mitigation 
in cellular systems with distributed components
such as heterogeneous 4G LTE systems that include small cells.
Small cells are basestations that aim to provide high data rate
coverage over a small high-traffic area. For example, {\em picocells} are
owned by a cellular provider and placed on public locations such as
lamp posts.  Alternatively, {\em femtocells} are owned by an end-user with
the aim of improving coverage in a private home or business.  An
important property of femtocells is that they can operate in Closed
Subscriber Group (CSG) mode in which the basestation restricts the set
of mobile terminals that can connect to it. Another effect introduced
by femtocells is that interference to macrocell users can now come
from a femtocell in the interior of the macrocell itself, not just
from neighboring macrocells.

LTE networks with small cells represent a hybrid of traditional
cellular networks and traditional ad-hoc networks.  On the one hand,
basestations are running the full LTE protocols which allows for
the many interference mitigation schemes that these protocols provide.
On the other hand, the placement of picocells and femtocells in an LTE
network is likely to be unstructured and so the interference
configurations are likely to resemble a typical ad-hoc configuration.
As a consequence, there is no hope for any centralized planning, which
is a possibility for cellular network interference mitigation
techniques such as frequency reuse.  We therefore need distributed
algorithms.


We are interested in scheduling algorithms for a heterogeneous system
that consists of a mixture of macrocells and small cells.
We wish to determine
the channels, or carriers, used by each basestation as well as the
transmission rate on each channel.  This should be done in order to
maximize a utility function associated with the system. Although a
number of scheduling algorithms have been proposed in the LTE context,
many of them require a non-trivial amount of signaling among the
transmitters. For example, in some algorithms a scheduling decision
is preceded by a calculation of how the decision would affect the
overall system utility, e.g.\ by exchanging partial derivative
information between neighboring transmitters.  This is difficult to
support in heterogeneous networks with small cells due to the
complexity of setting up the necessary communication channels.

Our main result is to show that LTE scheduling in heterogeneous networks can be performed using techniques developed in the context of 802.11 networks. 
These networks utilize a Carrier-Sense Multiple
Access (CSMA) protocol, often enhanced with a
Request-to-Send/Clear-to-Send (RTS/CTS) mechanism \cite{Karn90}. In this
setup, transmitters sense the channel before transmitting and 
proceed only if no conflicting transmissions
are active.  

The attractiveness of this framework is that CSMA scheduling
algorithms can achieve optimal throughputs without any explicit
signaling. Coordination is implicit in the ``collision'' mechanism
defined by the CSMA mechanism.  In particular, Jiang and Walrand
showed in \cite{JiangW08} that such mechanisms can be used to achieve
any set of feasible throughputs.  Among the sequence of papers that
followed, Liu et al.\cite{LiuYPCP10} presented utility-optimal
algorithms for CSMA networks with a single carrier and single
transmission rate.  Two subsequent papers discussed implementation issues associated with these algorithms~\cite{LeeYCNKC12} and presented an extension to the case of a single transmission rate on multiple carriers~\cite{ProutiereYLC10}.
However, to the best of our knowledge no previous work has looked at
how such techniques can be applied in the LTE context.


A number of issues arise if we are to use CSMA-based scheduling
algorithms for LTE.  First, LTE
networks utilize an OFDM physical layer which consists of 
multiple transmission rates over multiple carriers realized by multiple adjustable power levels.
Hence, in addition to deciding when to transmit on each carrier as in
\cite{JiangW08,LiuYPCP10,ProutiereYLC10}, the scheduler now also chooses 
the power level used and the transmission
rate. Second, LTE networks do not have an explicit carrier sense
mechanism to detect conflicting transmissions. We need to build this
capability via the existing Channel Quality Indication (CQI)
mechanism. Lastly, each basestation in an LTE network typically has
its own local scheduler, such as Proportional Fair, that governs the
transmissions to the users within the cell.  We need a mechanism
that allows 
the existing local scheduler to work with the CSMA-based algorithm.


The main result of this paper is a scheduling algorithm for utility
maximization in heterogeneous LTE networks. 
Our methodology is motivated by the CSMA analysis of \cite{LiuYPCP10}. 
We believe our main contribution is in showing that the analysis can be 
adapted for the case of multiple per-carrier transmission rates and powers, and 
(perhaps more importantly) describing how the algorithms can be implemented using the CQI mechanism present in LTE. 

We begin with a more
abstract version of the algorithm in which details of the interference
are abstracted away into a {\em feasibility region} for the
transmissions. Many of the existing CSMA algorithms implicitly work
with this notion.  We initially assume an ``oracle'' that informs a
transmitter whether a potential transmission would disrupt existing
transmissions. The later sections of the paper discuss how to realize
such a scheme in a heterogeneous LTE system in practice.
We structure the paper as follows.
\begin{itemize}
\item 
In Section \ref{s:abs} we present an abstract model of a multi-carrier system that allows for
multiple transmission
rates on each carrier. This model assumes that each transmitter knows
whether a potential transmission is feasible.
\item 
In Sections \ref{s:algo}-\ref{s:analysis} we adapt the Liu et al.\
single-carrier single-rate utility maximization algorithm for CSMA to
the abstract framework of Section~\ref{s:abs} to address the general
case of multiple transmission rates on multiple carriers. 
\item 
In Sections \ref{s:concrete} and \ref{s:imp} we give a concrete model for
heterogeneous LTE networks with small cells, in which we address power
level and interference directly instead of via the notion of a feasibility
region for transmission. 
We also discuss practical issues such as CQI-based
collision detection and incorporating a local scheduler.
\item 
In Section \ref{s:sim} we present simulation results.
\item 
In Section \ref{s:otherwork} we give an overview of past work on
scheduling and resource allocation in cellular and 802.11 networks.
\end{itemize}

\section{Abstract Model}
\label{s:abs}
We begin by describing an abstract model that captures the notion of
multiple transmission rates on multiple carriers and variable-power scheduling.  We consider a
system in which a set of transmitters communicate to a set of
receivers via a set of links $\cL$ on a set of carriers $\cC$ at
transmission rates from a set $\cR$ of positive numbers.  Each link is
associated with a transmitter and a receiver, where multiple links may
share a common transmitter but each link corresponds one-to-one with a
receiver.  

\cm{
For simplicity we assume links, i.e.\ the pairing of
transmitters and receivers, are given and fixed, as
scheduling decisions take place at a finer timescale than the
timescale of channel variations.
We
also assume that carriers from $\cC$ are sufficiently spaced that
a transmission on one carrier does not interfere with that on any
other carrier. 
It is easy to see
the correspondence between transmitters and
basestations, receivers and mobile users, links and downlink
transmissions, carriers and LTE resource blocks, and time slots and
TTIs (Transmission Time Intervals in LTE). 
}

We solve a scheduling problem, i.e.\ at each time instant we specify
for every carrier the links that are transmitting on that carrier
together with the associated transmission rates.  More precisely, we
represent a {\em schedule} on a carrier $c\in \cC$ by a vector
$(r_0,\ldots,r_{L-1})\in \cR_0^L$ where $l=|\cL|$ and
$\cR_0 = \cR \cup \{0\}$. Such a schedule is
{\em feasible} if it can be realized by an appropriate 
power allocation so that every link $\ell\in \cL$ can transmit from
its transmitter to its receiver at rate $r_{\ell}$ on carrier $c$ 
simultaneously.  A scheduling algorithm describes a schedule for
every carrier $c\in\cC$ at every time instant.  Note that a link is
allowed to simultaneously transmit on multiple carriers.

This abstract model captures interference and power assignments by the
notion of a {\em feasibility region}, which consists of all valid
schedules.  In the next two sections we describe our basic algorithmic
framework in this abstract model for which we do not concern ourselves
with how the system knows whether a schedule is feasible but simply
assume an {\em oracle} that indicates whether a potential transmission
leads to feasibility.  As mentioned before, for 802.11 networks this
can be approximately realized by CSMA techniques coupled with RTS/CTS
messages.  In later sections we describe how the oracle can be
realized in an LTE heterogeneous network.


We consider the problem of system utility maximization. In
particular let $\gamma_{\ell,c}$ indicate the transmission rate on
link $\ell$ on carrier $c$. For a given concave utility function
$U(\cdot)$ we wish to maximize the aggregate utility over all links, i.e.\
to maximize $\sum_{\ell} U(\sum_c \gamma_{\ell,c})$. Note that
for each link the utility function is applied to the total
transmission rate on the link over all carriers.  This coupling
between the carriers implies that we cannot simply treat each carrier
as an isolated system.


A formal version of the optimization problem is given below.
Consider a schedule $m\in \cN_c$ where $\cN_c$ is the set of feasible
schedules on carrier $c$. If
$\pi_m \in [0,1]$ indicates the fraction of time that $m$ occurs and $r_{\ell,m}$
indicates the transmission rate on $\ell$ under schedule $m$, then
$\gamma_{\ell,c}$ can be viewed as a weighted sum of $r_{\ell,m}$
where $\pi_m$ serves as the weight. Throughout the paper,
notations such as $\vec \gamma$
indicate a vector $(\gamma_{\ell,c})_{\ell\in \cL, c\in\cC}$.
We wish to solve:
\begin{eqnarray}
\max && f_1(\vec\gamma)=\sum_{\ell\in \cL} U(\sum_c \gamma_{\ell,c}) \label{eq:maxutil}
\\
\mbox{s.t.}&& \gamma_{\ell,c} \le \sum_{m\in \cN_c}
r_{\ell,m}\pi_{m}~~~\forall \ell,c \nonumber \\
&& \sum_{m\in \cN_c}\pi_{m}=1~~~\forall c. \nonumber
\end{eqnarray}
We can view this as an extension of the formulation of
\cite{ProutiereYLC10} that allows for multirate transmissions (and
implicitly variable transmission powers).

In reality the sets $\cL$ and $\cN_c$ can change over time due to 
mobility. However, as in common in the literature we assume that this
happens on a slow enough timescale that it makes sense to solve the
utility maximization problem for the current network configuration. 

\section{Multi-carrier, multi-rate scheduling algorithm}
\label{s:algo}
We first describe a routine {\sc RandAcc} (in Figure
Algorithm~\ref{alg:CSMA}), a continuous-time random access algorithm
that determines when a transmission will take place. We use
$\langle\ell,r,c\rangle$ to denote the transmission on link
$\ell\in\cL$ on carrier $c\in \cC$ at rate $r\in \cR$. Each
transmission $\langle \ell,r,c\rangle$ is associated with two
parameters, $\lambda_{\ell,r,c}$ the channel access rate and
$\mu_{\ell,r,c}$ which represents the expected transmission
duration. After an exponentially distributed waiting period with mean
$1/\lambda_{\ell,r,c}$,
%
%
{\sc RandAcc} checks whether $\langle \ell, r,c\rangle$ leads to a
valid schedule in $\cN_c$ at that time instant. If yes, the
transmission starts immediately and lasts for an exponentially
distributed time period with mean $\mu_{\ell,r,c}$. Note that an
invalid schedule includes the situation in which $\ell$ conflicts with
itself, namely $\ell$ already transmits on $c$, or $\ell$ conflicts
with another link on $c$, both of which are captured by $\cN_c$.  Note
also that since {\sc RandAcc} operates in a continuous manner, two
links make a scheduling decision simultaneously with zero probability
and therefore they make conflicting decisions with zero probability.


\begin{algorithm}[btp]
\caption{{\sc RandAcc}($\langle\ell,r,c\rangle$, $\lambda$, $\mu$, $T$)}
\label{alg:CSMA}

\begin{algorithmic}[20]
\STATE $\bullet$ $\ell:$ link
\STATE $\bullet$ $r:$ rate
\STATE $\bullet$ $c:$ carrier
\STATE $\bullet$ $\lambda:$ channel access rate
\STATE $\bullet$ $\mu:$ expected transmission duration
\STATE $\bullet$ $T:$ time frame
\STATE
\STATE $t \leftarrow$  beginning of $T$
\WHILE{$t\in T$}
\STATE $x\leftarrow$ randomly drawn from $Exp(\lambda)$
\STATE $t\leftarrow  t+ x$
\IF{$\langle \ell,r,c\rangle$ leads to config.\ in $\cN_c$}
\STATE $x\leftarrow$ randomly drawn from $Exp(1/\mu)$
\STATE $\ell$ transmits on $c$ at $r$ during $(t,t+x]$
\STATE $t\leftarrow t+x$
\ENDIF
\ENDWHILE
\end{algorithmic}
\end{algorithm}

The algorithm MMUO (multi-carrier multi-rate utility optimization)
approximates a solution to (\ref{eq:maxutil}) as follows (see
Algorithm \ref{alg:UO-CSMA}).  Time is divided into frames of fixed
duration. During each frame $f$, each potential
transmission $\langle\ell,r,c\rangle$ calls the {\sc
RandAcc} routine with parameters
$\lambda_{\ell,r,c}[f]$ and $\mu_{\ell,r,c}[f]$.

At the end of each frame, link $\ell$ calculates the service received
during the frame (denoted by $S_{\ell}[f]$), and updates a virtual
queue size parameter (denoted by $q_{\ell}[f]$) as follows.
\begin{eqnarray}
q_{\ell}[f+1] = \Big[q_\ell[f] 
+ b[f]\cdot
\Big(U'^{-1}(q_\ell[f]/V)-S_\ell[f]\Big)\Big]^{q_{\max}}_{q_{\min}} 
\label{eq:q}  
\end{eqnarray}
In the above equation, $b$ is a step size function that satisfies
property $(A1)$ (defined later), $q_{\max}$ and $q_{\min}$ are bounds
on the virtual queue size, and $[x]^{q_{\max}}_{q_{\min}} =
\min(q_{\max},\max(q_{\min},x))$.  The positive parameter $V$ controls
the accuracy of the algorithm.

The values of $\lambda_{\ell, r,c}[f]$ and $\mu_{\ell,
r,c}[f]$ stay unchanged during each frame $f$, and are updated to
$\lambda_{\ell, r,c}[f+1]$ and $\mu_{\ell, r,c}[f+1]$ at the end of
frame $f$, so that
\begin{eqnarray}
\lambda_{\ell, r,c}[f+1]\cdot \mu_{\ell, r,c}[f+1]
=\exp(r\cdot q_\ell[f+1]).
\label{eq:product}
\end{eqnarray}
As we shall see in (\ref{eq:pi}) the performance of {\sc RandAcc} depends
on the product of $\lambda$ and $\mu$. The choice of this product is
explained in the proof of Theorem~\ref{t:uo-csma}. 

\begin{algorithm}[btp]
\caption{MMUO}
\label{alg:UO-CSMA}

\begin{algorithmic}[10]
\FOR{each time frame $f$}
\FOR{each link $\ell$, carrier $c$ and rate $r$}
\STATE
{\sc RandAcc}($\langle \ell,r,c\rangle$, $\lambda_{\ell,r,c}[f],\mu_{\ell,r,c}[f]$, $f$)
\ENDFOR
\FOR{each $\ell$}
\STATE update service received $S_\ell[f]$
\STATE update virtual queue $q_\ell[f+1]$ according to (\ref{eq:q}) 
\ENDFOR
\FOR{each $\ell$, $c$ and $r$}
\STATE update $\lambda_{\ell,r,c}[f+1]$ and $\mu_{\ell,r,c}[f+1]$ 
according to (\ref{eq:product})
\ENDFOR
\ENDFOR
\end{algorithmic}
\end{algorithm}

\section{Analysis}
\label{s:analysis}
We now show that the MMUO algorithm leads to an optimal solution to
(\ref{eq:maxutil}). The proof builds upon the following optimality
properties of {\sc RandAcc} that were shown in \cite{JiangW08} for the special case
of one carrier and 0/1 transmission rates.

\subsection{{\sc{RandAcc}} for 0/1 Rates}
Suppose each link $\ell\in\cL$ calls {\sc RandAcc} with parameters
$(\lambda_\ell, \mu_\ell)$ over a sufficiently long time frame.  Let
$\vec \lambda = (\lambda_\ell)_{\ell\in\cL}$ and let $\vec\mu =
(\mu_\ell)_{\ell\in\cL}$. Let $m^{\vec\lambda, \vec\mu}(t)$ be the
schedule at time $t$ under the {\sc RandAcc} routine.  Recall
$r_{\ell,m}$ is the transmission rate on link $\ell$ under schedule
$m$. In this case $r_{\ell,m} \in\{ 0,1\}$.

\begin{lemma}[\cite{Kelly79}]
\label{l:kelly}
The sequence of schedules $m^{\vec\lambda,\vec\mu}(t)$ for $t\ge 0$
form a continuous time reversible Markov Chain with the following
stationary distribution.
\begin{eqnarray*}
\pi_m^{\vec\lambda_c,\vec\mu_c} =
\frac{\Pi_{\ell:r_{\ell,m}=1}\lambda_{\ell}\cdot\mu_{\ell}}
{\sum_{m'} \Pi_{\ell: r_{\ell,m'}=1}\lambda_{\ell}
\cdot\mu_{\ell}} &\forall m,
\end{eqnarray*}
where $\Pi_{\ell\in\emptyset}(\cdot)=1$.
\end{lemma}
Further, the resulting link throughput
$\gamma_{\ell}^{\vec\lambda,\vec\mu} = \sum_{m}
\pi_m^{\vec\lambda_c,\vec\mu_c}r_{\ell,m}$ is optimal for every link
$\ell\in\cL$ in the following sense.  We say that a link throughput
vector $\vec \gamma$ is feasible if there exist $\pi_m$ such that
$\gamma_{\ell}<\sum_{m\in\cN_c} r_{l,m}\pi_m$. 
\begin{lemma}[\cite{JiangW08}]
\label{l:jiang}
For any feasible link throughput vector $\vec \gamma$,
there exists $\vec\lambda$ and $\vec\mu$ such that
$$
\gamma_\ell \le \gamma_\ell^{\vec\lambda,\vec\mu}.
$$
\end{lemma}

\subsection{MMUO}
For multiple transmission rates on multiple carriers, the generalization from
Lemma~\ref{l:kelly} to Corollary~\ref{c:kelly} is straight-forward.
This is because {\sc RandAcc} that runs on one carrier does not interfere
with that on a different carrier. Further, each rate-link pair can be
treated as a distinct link. For a particular carrier $c$, let
vectors $\vec\lambda_c= (\lambda_{\ell,r,c})_{\ell\in\cL, r\in \cR}$
and $\vec\mu_c= (\mu_{\ell,r,c})_{\ell\in\cL, r\in \cR}$.
Recall $\cN_c$ is the set of feasible schedules on carrier $c$.
\begin{cor}
\label{c:kelly}
The schedule sequence $m^{\vec\lambda_c,\vec\mu_c}(t)$ for $t\ge 0$
is a continuous time reversible Markov Chain with the following
stationary distribution.
\begin{eqnarray}
\pi_m^{\vec\lambda_c,\vec\mu_c} =
\frac{\Pi_{\ell:r_{\ell,m}>0}\lambda_{\ell,r_{\ell,m},c}\cdot\mu_{\ell,r_{\ell,m},c}}
{\sum_{n\in \cN_c} \Pi_{\ell: r_{\ell,n}>0}\lambda_{\ell,r_{\ell,n},c}
\cdot\mu_{\ell,r_{\ell,n},c}} &
\forall m\in\cN_c \label{eq:pi} 
\end{eqnarray}
\end{cor}

The above expression shows that the stationary distribution only
depends on the product of the  parameters $\lambda$ and $\mu$.
Whenever MMUO invokes {\sc RandAcc}, it does so with $\lambda$ and $\mu$
parameters that are set according to (\ref{eq:product}).  For any vector of
virtual queues $\vec q=(q_{\ell})_{\ell\in\cL}$, we denote by $\pi^{\vec
q}$ the resulting distribution on $\cN_c$ from the {\sc RandAcc} routine.
From (\ref{eq:pi}) we have,
\begin{eqnarray}
\pi_m^{\vec q} =
\frac{\exp\left(\sum_{\ell:r_{\ell,m}>0}r_{\ell,m}\cdot q_\ell\right)}
{\sum_{m'\in \cN_c} \exp
\left(\sum_{\ell:r_{\ell,m'}>0}r_{\ell,m'}\cdot q_\ell\right)}
&
\forall m\in\cN_c \label{eq:pi2} 
\end{eqnarray}
The resulting link throughput is therefore,
\begin{eqnarray}
\gamma_{\ell,c}^{\vec q}  =  
\sum_{m\in \cN_c} \pi_m^{\vec q} r_{\ell,m}
& \forall \ell, c
\label{eq:gamma2}
\end{eqnarray}
For utility optimization, our goal is to show that the virtual queues
under MMUO converge to a vector $\vec q^*$ so that the above link
throughput under $\vec q^*$ maximizes the utility as defined
in~(\ref{eq:maxutil}).  
Note that for this problem we cannot treat each carrier in isolation since the throughput of a link is aggregated over all carriers. 
Note also that the 
optimization problem (\ref{eq:maxutil2}) in Theorem~\ref{t:uo-csma} 
differs from that of~(\ref{eq:maxutil}) in its objective function,
but shares the same constraints. The motivation of this
reformulation is to obtain a more useful set of KKT conditions. 
Further, Theorem~\ref{t:uo-csma} will state that the
optimal values of the two objective functions can be arbitrarily close.
The main result of this section is captured in the following theorem. 
It relies on two standard but technical assumptions
$(A1)$ and $(A2)$ which we detail below. 
\begin{theorem}
Under assumptions $(A1)$ and $(A2)$, for any initial condition $\vec q[0]$,
MMUO converges in the following sense.
\begin{eqnarray*}
\lim_{f\leftarrow \infty} \vec q[f]=\vec q^* 
\end{eqnarray*}
where $\vec q^*$ and $\vec\gamma^*$ are such that 
$(\vec\gamma^*,\vec\pi^{\vec q^*})$
is the solution to the following convex optimization problem over
$\vec\gamma$ and $\vec\pi$.
\begin{eqnarray}
\max && 
f_2(\vec\gamma, \vec \pi)= \nonumber\\
&&
V\sum_{\ell\in \cL} U(\sum_{c} \gamma_{\ell,c}) - \sum_c
\sum_{m\in \cN_c} \pi_m\log \pi_m \label{eq:maxutil2}\\
\mbox{\em{s.t.}} && \gamma_{\ell,c} \le \sum_{m\in \cN_c}
r_{\ell,m}\pi_{m}~~~\forall \ell,c \nonumber \\
&& \sum_{m\in \cN_c}\pi_{m}=1~~~\forall c. \nonumber
\end{eqnarray}
Further, if $\vec \gamma^\dagger$ is the optimal solution
to (\ref{eq:maxutil}), then
\begin{eqnarray}
|f_1(\vec \gamma^*)-f_1(\vec \gamma^\dagger)|
 \le |\cC|\log|\cup_c\cN_c| /V
\label{eq:diff}
\end{eqnarray}
\label{t:uo-csma}
\end{theorem}

The assumptions of Theorem~\ref{t:uo-csma} are:
\begin{enumerate}
\item[$(A1)$]
$\sum_{f=0}^{\infty} b[f]=\infty$ and
$\sum_{f=0}^{\infty} b^2[f] <\infty$.
\item[$(A2)$]
If $\vec p^o\in \Re_+^L$ is a solution to 
\begin{eqnarray*}
U'^{-1}(p_{\ell}/V)-\sum_c\sum_{m\in \cN_c}r_{\ell,m}\pi_m^{\vec p} =0 
&&
\forall \ell \in \cL
\end{eqnarray*}
then $q_{\min}\le p^o_\ell \le q_{\max}$ for
all $\ell\in\cL$.
\end{enumerate}
The parameters $b[\cdot]$ will only be used in the analysis, not in the algorithm itself. In addition, the parameters $q_{\min}$ and $q_{\max}$ are under our control. Hence for any problem instance we can make sure that Assumptions (A1) and (A2) hold. 

\begin{proof}
There are several steps in the proof. In the first two steps we follow
the framework of \cite{LiuYPCP10} to show that the dynamics of MMUO
can be captured by a system of differential equations. In the third
step we must deviate from the approach of \cite{LiuYPCP10} in order to
handle multiple transmission rates on multiple carriers. 

We begin by replacing the discrete time frames of MMUO with a more convenient continuous interpolation.
For notation we use square brackets $[\cdot]$ indexed with
integers $f$ for discrete sequences defined on frames $f$, and we use
round brackets $(\cdot)$ indexed with real numbers $t$ for a continuous
{\em scaled} version of time.  For the discrete time sequence of virtual queue
vectors $\vec q[f]$ for integral frames $f=1,2, \dots$, we define as follows
its continuous interpolation $\vec q(t)$ for all real positive numbers
$t$. We also define a continuous version $S_\ell(t)$ for the discrete
service sequence $S_\ell[f]$.  For $f=1,2,\dots$, let $t_f =
\sum_{i=1}^f b[i]$. For $t\in [t_f, t_{f+1})$, let
\begin{eqnarray*}
q_{\ell}(t)&=&q_{\ell}[f]\cdot{{t_{n+1}-t}\over{t_{n+1}-t_n}}+
q_{\ell}[f+1] \cdot{{t-t_n}\over{t_{n+1}-t_n}} \\
S_{\ell}(t) &=& S_\ell[f]
\end{eqnarray*}
(In other words the continuous time process $t$ is created from the discrete time frames scaled by the intervals $[t_f,t_{f+1})$.)

The following lemma is shown in \cite{LiuYPCP10} and says that the
continuous sequence $\vec q(t)$ converges to the solution of the system of
stochastic differential equations~(\ref{eq:deq1}) which can be viewed as a
continuous version of (\ref{eq:q}).  The equations are stochastic due to the 
term $S_\ell(t)$ which is the result of the stochastic process of
MMUO.
\begin{lemma}[\cite{LiuYPCP10}]
\label{l:continuous}
Let $\vec p^*$ be the solution to the following system of differential
equations with variable $\vec p=(p)_{\ell\in\cL}$.
\begin{eqnarray}
\dot p_{\ell} = 
\Big[U'^{-1}(p_{\ell}/V) -S_{\ell}(t)\Big] \cdot 
1_{p_{\ell}\in [q_{\min},q_{\max}]},
\label{eq:deq1}
\end{eqnarray}
where $1_{p_{\ell}\in [q_{\min},q_{\max}]}$ is an indicator variable
for whether $p_{\ell}$ is in the range of $[q_{\min},q_{\max}]$.  Fix
any time instant $\tau$. If $\vec p^*(\tau)=\vec q(\tau)$, then
$\lim_{\tau\rightarrow\infty}\sup_{t\in[\tau,\tau+T]}\| \vec p^*(t)-\vec
q(t)\| =0$.
\end{lemma}

The second step of the proof says that any fixed point of the
stochastic system (\ref{eq:deq1}) will also be a fixed point of an
associated deterministic system of equations.  Note that the {\sc
RandAcc} routine may not converge to the stationary distribution of
(\ref{eq:pi}), or equivalently (\ref{eq:pi2}), with the given
$\lambda$ and $\mu$ parameters within each frame before the parameters
are updated to their respective new values for the next frame.  Hence
the service $S_{\ell}(t)$ in the above differential equations is a
stochastic quantity.  In other words, when MMUO invokes the {\sc
RandAcc} routine, if each frame $f$ is sufficiently long,
$S_{\ell}(t)$ would converge to its long-term average
$\sum_c\sum_{m\in \cN_c}r_{\ell,m}\cdot
\pi_m^{\vec q}$ and we could replace the stochastic term $S_{\ell}(t)$
in (\ref{eq:deq1}) by its long term average. This would result in
the following system (\ref{eq:deq2}). Unfortunately, in reality, the frame $f$ may not be long
enough.  However, the following result still allows us to say that
systems (\ref{eq:deq1}) and (\ref{eq:deq2}) are closely related,
regardless of the convergence of {\sc RandAcc}. As we shall see,
(\ref{eq:deq2}) also provides a connection to solving
(\ref{eq:maxutil2}) via the KKT conditions.
\begin{lemma}[\cite{LiuYPCP10}]
\label{l:converge}
Every limit point of system (\ref{eq:deq1}) is almost always a fixed
point of the following system.
\begin{eqnarray}
\dot p_{\ell} =
\Big[U'^{-1}(p_{\ell}/V) -\sum_c\sum_{m\in \cN_c}
r_{\ell,m}\cdot \pi_m^{\vec p} \Big] \cdot 
1_{p_{\ell}\in [q_{\min},q_{\max}]}, \nonumber \\
\forall \ell\in\cL,
\label{eq:deq2}
\end{eqnarray}
where $\pi_m^{\vec p}$ is defined in (\ref{eq:pi2}).
\end{lemma}

In the third step of the proof we show that the system (\ref{eq:deq2})
leads to a solution of the optimization problem
(\ref{eq:maxutil2}). (It effectively solves the problem via the
subgradient method). For this step we must deviate from the analysis
of \cite{LiuYPCP10} in order to handle the multi-carrier and multi-rate
aspects of (\ref{eq:maxutil2}).  In particular, the Lagrangian of
(\ref{eq:maxutil2}) is given by:
\begin{eqnarray*}
&& L(\vec \gamma,\vec \pi:\vec \nu,\vec \eta) \\
&=& 
\sum_{\ell\in \cL} \left(V\cdot U(\sum_{c} \gamma_{\ell,c}) - 
\sum_c \nu_{\ell,c}\gamma_{\ell,c}\right)\\
&+& \sum_c \left(\sum_{\ell\in \cL} \nu_{\ell,c} 
\sum_{m\in \cN_c} r_{\ell,m}\pi_{m} -
\sum_{m\in \cN_c}\pi_{m}\log \pi_{m}\right)\\
&-& \sum_c \eta_c \left(\sum_{m\in \cN_c}\pi_{m}-1\right).
\end{eqnarray*}.

Hence the KKT conditions for (\ref{eq:maxutil2}) are:
\begin{eqnarray}
VU'(\sum_c \gamma_{\ell,c}) = \nu_{\ell,c}&& 
\forall \ell, c \label{eq:kkt1} \\
- 1-\log \pi_{m} + \sum_{\ell}r_{\ell,m}\nu_{\ell,c}- \eta_{c} = 0
&& \forall m\in \cN_c, \forall c \label{eq:kkt2}\\
\gamma_{\ell,c} \le \sum_{m\in \cN_c}
r_{\ell,m}\pi_{m} && \forall \ell, c \label{eq:kkt3} \\
\nu_{\ell,c}\times (\gamma_{\ell,c}-\sum_{m\in \cN_c}r_{\ell,m}\pi_{m})=0 
&& \forall \ell, c  \label{eq:kkt4} \\
\nu_{\ell,c}\ge 0 && \forall \ell, c \label{eq:kkt5}\\
\sum_{m\in \cN_c}\pi_{m}-1=0&& \forall c\label{eq:kkt6} 
\end{eqnarray}
Inequalities (\ref{eq:kkt1}) and (\ref{eq:kkt2}) state the gradient
of the Lagrangian
$L(\vec \gamma,\vec \pi:\vec \nu,\vec \eta)$ is zero with respect
to the variables $\vec \gamma$ and $\vec \pi$. 
Inequalities (\ref{eq:kkt3}), (\ref{eq:kkt4}) and (\ref{eq:kkt5})
state that the first set of constraints of (\ref{eq:maxutil2})
hold and has zero duality gap. The last equality (\ref{eq:kkt6})
states the second set of constraints of (\ref{eq:maxutil2}) hold. 

We introduce a new variable $\vec p = (p)_{\ell\in\cL}$ and set the primary variables via
\begin{eqnarray*}
\pi_m  =  \pi_m^{\vec p} \mbox{ as in (\ref{eq:pi2})} && \forall m
\end{eqnarray*}
and the dual variables via
\begin{eqnarray*}
\nu_{\ell,c}&=&p_{\ell} ~~~~~~~ \forall \ell, c\\
\eta_c &=& \log\left(\sum_{m\in \cN_c}\exp(\sum_{\ell} 
r_{\ell,m}q_{\ell})\right)-1 ~~~~~~ \forall c.
\end{eqnarray*}
We can see that the KKT conditions (\ref{eq:kkt2}), 
(\ref{eq:kkt5}) and (\ref{eq:kkt6}) are easily satisfied due to the
definition of $\pi_m^{\vec p}$ in (\ref{eq:pi2}) and as long as
$\vec p \in\Re_+^L$.

We next aim to satisfy the remaining KKT conditions (\ref{eq:kkt1}),
(\ref{eq:kkt3}) and (\ref{eq:kkt4}) via the subgradient method. From (\ref{eq:kkt1}),
(\ref{eq:kkt4}) and the fact that $\nu_{\ell,c}$ is set to $p_\ell$,
we wish to have,
\begin{eqnarray*}
p_\ell \times (U'^{-1}(p_\ell/V)
-\sum_c\sum_{m\in \cN_c}r_{\ell,m}\pi_m^{\vec p}) = 0. &&\forall \ell
\end{eqnarray*}
If this does not hold (and hence we are not yet at optimality), the
standard subgradient method updates the $p_{\ell}$ according to the
following system of differential equations.
\begin{eqnarray}
\label{eq:deq3}
\dot{p}_{\ell} = 
U'^{-1}(p_{\ell}/V)-\sum_c\sum_{m\in \cN_c}r_{\ell,m}\pi_m^{\vec p}.
\end{eqnarray}
Due to the convexity of the problem (\ref{eq:maxutil2}) the system (\ref{eq:deq3}) will eventually converge to a fixed point, $\vec p^*$.
Note that the system (\ref{eq:deq3}) is identical 
to (\ref{eq:deq2}) as long as $p_\ell\in [q_{\min},q_{\max}]$
for all $\ell\in \cL$. Due to assumption ($A2$) and the definition of
$\nu_{\ell,c}$, the solution to the dual of (\ref{eq:maxutil2})
without the constraint $p_\ell\in [q_{\min},q_{\max}]$ falls in the
range of $[q_{\min},q_{\max}]$ and is therefore equivalent to the fixed point $\vec p^*$ of the system
(\ref{eq:deq3}).

With $\vec p^*$ chosen, KKT condition (\ref{eq:kkt1}) is satisfied.
We now set 
\begin{eqnarray*}
\gamma_{\ell,c}  =  
\sum_{m\in \cN_c} \pi_m^{\vec p^*} r_{\ell,m}
&& \forall \ell, c,
\end{eqnarray*}
which satisfy the remaining conditions of (\ref{eq:kkt3}) and (\ref{eq:kkt4}).

Finally, since
\begin{eqnarray*}
f_2(\vec \gamma^*, \vec \pi^*)
&\ge& f_2(\vec \gamma^\dagger, \vec \pi^\dagger)\\
f_1(\vec \gamma^*)
&\le& f_1(\vec \gamma^\dagger)
\end{eqnarray*}
and the entropy $\sum_m \pi_m\log\pi_m\le \log|\cup_c\cN_c|$, the
proof of Theorem~\ref{t:uo-csma} is complete.
\end{proof}
We conclude by briefly summarizing in what sense we have
shown that MMUO is optimal. We have shown that an appropriate
continuous interpolation of the virtual queue dynamics approaches in
the limit a vector $\vec p^*$ that defines an optimal dual solution of
(\ref{eq:maxutil2}) (via the KKT conditions). Via (\ref{eq:product})
the optimal virtual queue sizes determine channel access parameters
$\lambda_{\ell,r,c}$ and $\mu_{\ell,r,c}$ for which the corresponding
link throughputs provide an optimal primal solution to problem
(\ref{eq:maxutil2}). 

\section{A More Concrete Model: Heterogeneous LTE System}
\label{s:concrete}

In this section we present a more concrete model for the scheduling
problem so that it more closely matches resource allocation in LTE
heterogeneous networks. 

We begin with a brief system description. We consider downlink
transmissions from a set of basestations to a set of mobile users in a
time-slotted system.  We assume an OFDM-based air interface in which
the spectrum is divided into a set of carriers called resource blocks
(RBs), each of which can be scheduled separately. For example a 20MHz
LTE system is typically divided into 100 resource blocks.  In the time
dimension a time slot corresponds to a Transmission Time Interval
(TTI) which has a typical duration of 1ms in an LTE system.

We consider a heterogeneous network in which the basestations are
divided into two classes, namely macrocells and femtocells.  (For ease
of description we use the terms ``macros'' and ``femtos''. However,
our discussions also apply directly to networks with picocells.)
Macros typically have a much higher max transmit power than femtos,
since macrocells provide wide-area coverage, whereas femtocells (which
may be privately owned) provide focused coverage in one specific
location, e.g.\ a house or apartment.
At any time instant each mobile user associates with one
basestation. Each macro accepts an association with any mobile user. A
femto however may be in ``Closed Subscriber Group'' mode (CSG) and
only accept an association with a small subset of users.  We remark
that femtos have two notable effects that are departures from
traditional cellular networks.  First, they may create strong
interference to a macrocell from within the cell itself, whereas in a
macro-only network interference to a cell mostly comes from outside
that cell.  Second, a mobile user may not be able to associate with
the basestation with the strongest signal if the basestation is a
femto in CSG mode and cannot associate with the user.  We assume that
each mobile user associates with the basestation for which the
received signal is strongest, among those that the user is able to
associate with.

Unlike in the abstract model, we address basestation power allocation and
interference directly instead of via the notion of feasibility
regions.  For basestation $i$, let $U_i$ be the set of associated
users. To abuse notation, we also use $U_i$ to denote the set of
links that are incident to $i$ as there is a one-to-one correspondence
between the users and links.

The maximum transmit power $p_i$ for basestation $i$ is given and
fixed. The scheduling problem is how to distribute $p_i$ among the
resource blocks $c\in \cC$ and among the users in $U_i$.  Let
$p_{i,c}(t)$ be the power allocation of $p_i$ on resource block $c$ at
time $t$; let $p_{i,c,j}(t)$ be the allocation of $p_{i,c}(t)$ on user
$j\in U_i$.  Note $\sum_c p_{i,c}(t)\le p_i$, and $\left\{
\begin{array}{llll} p_{i,c,j}(t)& =& p_{i,c}(t)& \mbox{ for one } j\in
U_i\\ p_{i,c,j'}(t)&=& 0 & \mbox{ for }j'\ne j
\end{array}\right..$
That is, $p_{i,c}$ is allocated entirely to {\em one} chosen user
$j\in U_i$.

Power settings and transmission rates are related through the 
{\em channel quality information} (CQI).  CQI values are
defined on pairs of links and resource blocks.  During every time slot
$t$, the values of $\mbox{CQI}_{c,\ell}(t)$ for all $c\in \cC$ and $\ell\in
U_i$ are reported to basestation $i$.  We assume that each basestation
has perfect CQI reporting.

Let $r_{c,\ell}(t)$ be the transmission rate along link $\ell$ on
carrier $c$ during time slot $t$. Specifically, for link $\ell=ij$ 
between the basestation $i$ and the associated user $j$, we define
\begin{eqnarray}
r_{c,\ell}(t) &=& w_c\cdot F(p_{i,c,j}(t)\cdot \mbox{CQI}_{c,\ell}(t)) 
\label{eq:ricl}\\
\mbox{CQI}_{c,\ell}(t) &=& {{g_{ijc}(t)}\over {
N_c + \sum_{i'\neq i}p_{i'c}g_{i'jc}(t)}} \label{eq:cqi}
\end{eqnarray}
In (\ref{eq:cqi}), $g_{ijc}$ represents the path loss between $i$ and
$j$ on resource block $c$, and $N_c$ is the background noise on
$c$. Both $g_{ijc}$ and $N_c$ depend on $c$ since radio propagation
conditions and background interference may be different on different
frequencies.  The product of $p_{i,c,j}$ and $\mbox{CQI}_{c,\ell}$ is
commonly referred to as {\em signal-to-interference-plus-noise ratio},
SINR. (We can therefore think of $\mbox{CQI}_{c,\ell}$ as the SINR for
a unit power transmission.)  In (\ref{eq:ricl}), $w_c$ is the
bandwidth of resource block $c$ and $F(\cdot)$ represents spectral
efficiency as a function of SINR.  For example $F(\cdot)$ could be a
suitably discretized version of the Shannon function $\log(1+x)$.  We
assume that $F(\cdot)$ is such that $r_{c,\ell}$ is always a member of
a discrete set $\cR\cup \{0\}$.

The primary scheduling decision is to determine the power levels
$p_{i,c}$. In the literature this problem is sometimes known as
inter-cell interference coordination (ICIC). The secondary scheduling
decision is to allocate $p_{i,c}$ to the user-level power $p_{i,c,j}$.
Typically, each basestation in an LTE network has its own local
scheduler for user-level allocation, in which case the scheduling
freedom is at the inter-cell level.  For concreteness we assume the
local scheduler uses the Proportional Fair (PF) algorithm.  In the
following section we give details on how to allocate $p_{i,c}$ and
$p_{i,c,j}$.  For user-level allocation $p_{i,c,j}$, we consider two
cases depending on whether a local scheduler exists.

\cm{
  Given the power setting, the value
of $r_{c,\ell}(t)$ is computed by applying a spectral efficiency
function $F(\cdot)$ to SINR$_{c,\ell}$, the
signal-to-interference-plus-noise ratio along link $\ell$ on resource
block $c$.
\begin{eqnarray}
r_{c,\ell}(t) &=& w_c\cdot F(\mbox{SINR}_{c,\ell}(t)) \label{eq:ricl}\\
\mbox{SINR}_{c,\ell}(t)&=&
p_{i,c,j}(t)\cdot {{g_{ijc}(t)}\over {
N_c + \sum_{i'\neq i}p_{i'c}g_{i'jc}(t)}} \mbox{~~~~~~~~~~}\nonumber \\
&& \hspace{1.2in}\mbox{ for link } \ell=ij\label{eq:sinr}.
\end{eqnarray}
Here, $w_c$ is the bandwidth of resource block $c$, link $\ell=ij$
is the link between the basestation $i$ and the associated user $j$,
$g_{ijc}$ represents the path loss between $i$ and $j$ on resource block $c$, and $N_c$ is
the background noise on resource block $c$. Both of the latter two quantities may depend on $c$ since radio propagation conditions and background interference may be different on different frequencies. 
We assume that $F(\cdot)$ is such that
$r_{c,\ell}$ is always a member of a discrete set $\cR\cup \{0\}$.

The second factor in the definition of SINR$_{c,\ell}(t)$ is 
called the channel quality information (CQI). Specifically,
\begin{eqnarray}
\mbox{CQI}_{c,\ell}(t) = {{g_{ijc}(t)}\over {
N_c + \sum_{i'\neq i}p_{i'c}g_{i'jc}(t)}}, \mbox{ for link } \ell=ij
\label{eq:cqi}
\end{eqnarray}
is reported to basestation $i$ during each time slot $t$. We
assume that each basestation has perfect CQI reporting.
}

\section{Implementation}
\label{s:imp}
In Section~\ref{s:algo} we described the utility-optimal MMUO
algorithm for the abstract model.  In this section we present
MMUO-based heuristics for the LTE resource allocation problem in
heterogeneous networks.  We address a number of issues. First,
scheduling decisions need to be made in slotted time rather than in
continuous time as in MMUO.  Second, scheduling decisions are about
setting power levels rather than transmission rates as in MMUO.
Third, we discuss how to incorporate a local scheduler such as
Proportional Fair.  Fourth, perhaps most significantly, we show
CQI-based methods for feasibility detection. This replaces the
feasibility oracle and the CSMA collision detection mechanism. Lastly,
since the basestations are divided into two classes, macro and femto,
interference can be reduced by not having every basestation compete on
every resource block.
 
We begin with a basic heuristic that bypasses the last two issues.  We
then describe three methods through which feasibility can be detected
in practice. We conclude with a modified heuristic in which macros and
femtos have priority on different sets of resource blocks.

\subsection{Basic Heuristic}
Our basic heuristic works very much in the spirit of MMUO.  To address
the first issue regarding slotted time each frame now consists of an
integral number of time slots. When the subroutine {\sc RandAcc} is
called with parameters $\lambda$ and $\mu$, the time between
transmission attempts (resp. the transmission period) is drawn from a
geometric distribution with mean $1/\lambda$ (resp. mean $\mu$).  One
problem is that two links may make decisions during the same time
slot. We can set the $1/\lambda$ values large enough so that this
rarely happens. If this does happen we assume that both conflicting
transmissions cease. A detailed explanation of how rare collisions
affect the performance of utility-optimal CSMA was given in
\cite{LiuYPCP10} and we can apply a similar analysis to MMUO. 

The output of MMUO, as described above, specifies the transmission
rate $r_{c,\ell}(t)$ for every transmission $\langle \ell,r,c\rangle$
that takes place during time slot $t$.  To obtain power settings,
equation (\ref{eq:ricl}) provides the direct
translation from transmission rates to power levels.
\begin{eqnarray}
p_{i,c,j}(t) = {{F^{-1}(r_{c,\ell}(t)/w_c)}\over {\mbox{CQI}_{c,\ell}(t)}},
\mbox{~~~~~~~~ for link } \ell=ij.
\label{eq:picj}
\end{eqnarray}
For a given basestation $i$ and resource block $c$, the feasibility
oracle guarantees that one user $j\in U_i$ has positive power
allocation $p_{i,c,j}(t)$.  Let $p_{i,c}(t)= p_{i,c,j}(t)$ for this
user $j$. The transmission now takes place as long as $\sum_c p_{i,c}(t)\le p_i$. 
This addresses the second issue.

If it is the case that we can specify both user-level as well as
inter-cell power allocations, we are done. However, as discussed in
Section~\ref{s:concrete}, in many instances we only have the freedom
for specifying $p_{i,c}$ since the user-level power is determined by a
local scheduler such as the Proportional Fair (PF) algorithm.  In this
case we run the MMUO algorithm ``in the background'' to compute the
$p_{i,c}$ values and then determine which user in $U_i$ receives the
transmission power $p_{i,c}$ using the PF algorithm. (We remark that
once the power levels are set then which user is chosen by PF does not
affect the interference experienced in other cells.)

The PF algorithm works as follows. For each link $\ell\in U_i$ basestation $i$
maintains an estimate $R_{\ell}$ of the recent average transmit rate
on link $\ell$, and allocates power $p_{i,c}$ exclusively to the link
$\ell$ that maximizes the ratio $\tilde r_{c,\ell}(t)/R_\ell$, where
$\tilde r_{c,\ell}(t)$ is the nominal rate if user $j$ has power
allocation $p_{i,c,j}(t)=p_{i,c}(t)$. Again from
(\ref{eq:ricl}), we have
\begin{eqnarray}
\tilde r_{c,\ell}(t) = w_c \cdot F\left(
p_{i,c,j}(t)\cdot \mbox{CQI}_{c,\ell}(t)\right)
\mbox{~~~~~for link }\ell=ij \label{eq:tilder}
\end{eqnarray}
After each scheduling decision $R_{\ell}$ is updated for each link
according to an exponential filter. This addresses the third issue.

\subsection{Methods for Implementing the Feasibility Oracle} 
We now examine options for the only part of the algorithm that
requires coordination among basestations, namely feasibility
detection. Since resource block power assignment is typically done on
a slower timescale than individual time slots, we are interested in
determining whether a set of transmissions will be feasible over
multiple timeslots. In particular, we do not want to declare a transmission feasible if this is only true for a single timeslot due to fast fading. 

We discuss multiple mechanisms which all use techniques that have been
proposed in the standardization process for heterogeneous
networks (e.g.~\cite{3GPP36921}). Our initial mechanisms use the
existing CQI channel with one extra piece of information which we call
the {\em activity indicator}.  We also allow for a basestation to
``overhear'' a link to which it is not associated.  Our later
mechanisms show how the algorithm could be implemented if we indeed
have a channel for exchanging information between basestations (such
as the X2 channel that is defined in LTE). The bit-rate of such channels is typically
limited and so we stress that all we need to exchange are short messages
such as the activity values. No detailed exchange of channel
state is required.


\paragraph*{Method 1} 
In this method, an activity indicator is reported along with the
CQI. Specifically, let $y_{\ell,r,c}(t)$ be the binary activity indicator
that is set to one if and only if MMUO makes a transmission $\langle
\ell,r,c\rangle$ during time slot $t$.  When
CQI$_{c,\ell}(t)$ is reported to basestation $i$ for which $\ell\in
U_i$, $y_{\ell,r,c}(t)$ is also reported if it is set to 1.  Each
basestation $i$ listens to all CQI that it can decode, not just the
CQI for links in $U_i$.  If $i$ hears $y_{\ell',r',c}=1$ for some
$\ell'$ on resource block $c$, then every potential transmission
$\langle \ell,r,c\rangle$, for $\ell\in U_i$ and $r\in \cR$, is
declared infeasible.  Note that this method is similar in spirit to
the Clear-to-Send (CTS) mechanism for 802.11. 


%

\paragraph*{Method 2} This method is less stringent than  Method
$1$ in declaring infeasibility. For each activity indicator
$y_{\ell,r,c}(t)=1$ we define the {\em safety margin} to be the ratio
between the currently achievable transmission rate $\tilde
r_{c,\ell}(t)$ and the actual rate $r_{c,\ell}(t)$ that is used by
MMUO.  This achievable rate can be computed from the 
CQI$_{c,\ell}(t)$ values together with the current power levels. We assume that the 
safety margins are transmitted on the CQI channel along with the activity indicators. 
For some threshold $\upsilon>1$
we say that the activity indicator is {\em safe} if the safety margin
is above $\upsilon$, {\em vulnerable} if the margin is between $1$ and
$\upsilon$, and {\em in outage} if the margin is below $1$. (Note that
if we are in outage then user $\ell$ could not receive data at rate
$r$ for the current CQI values.)  Method 2 is the same as method 1
except that basestation $i$ does not refrain from declaring a
potential transmission on $\ell\in U_i$ on block $c$ feasible, even if
it overhears an activity indicator $y_{\ell',r',c}(t)=1$ as long as
this indicator is currently safe.


The exact value of $\upsilon$ could be a network-wide configured parameter. Alternatively each basestation could gradually lower a local estimate of $\upsilon$ until it observes links going into outage. 


\paragraph*{Method 3} 
This method applies probing to feasibility detection. Whenever a
basestation $i$ needs to decide if MMUO could transmit on
$\langle\ell,r,c\rangle$, for $\ell\in U_i$, it briefly sets power
level $p_{i,c}$ on resource block $c$ and observes the effects on
other users. Here $p_{i,c}$ is the power necessary to carry out the
transmission $\langle\ell,r,c\rangle$ and can be calculated as in
(\ref{eq:picj}).  If basestation $i$ overhears that any activity
indicator moves into outage then it sets $p_{i,c}$ back to $0$ and
declares $\langle\ell,r,c\rangle$ infeasible. This method has the
drawback that it could send neighboring users into outage for short
periods (and this would need to rectified by more robust channel
coding on the data channels). However, it has the advantage that
basestation $i$ gets a much better sense of the ``damage'' that might
be caused by setting a particular power level $p_{i,c}$ on resource
block $c$.

\paragraph*{Methods 4-6} 
The next three methods are essentially the same as Methods
1-3. However, instead of basestations overhearing activity indicators
and their associated safety status, each basestation would directly
communicate their own activity indicators and safety margins to all
their neighboring basestations.  This can be done using a channel such
as the X2 channel in LTE that provides communication between neighboring basestations.
Note that this is a lightweight communication since the activity
indicator only has 2 possible values and the safety status has only 3
possible values. In particular the basestations would not be
exchanging any detailed channel state information.

\paragraph*{Relationship to current LTE proposals} 
We now briefly discuss how the above methods could fit with mechanisms
that have been proposed in LTE standards for interference
coordination.  In the document \cite{3GPP36921} on RF requirements for femtos,
three options are proposed for communication between macros and
femtos. The first is direct over-the-air communication. The second is
over-the-air via ``victim'' users. This would correspond to the
overhearing methods 1-3 proposed above in that the victim user
broadcasts channel quality information that indicates to an interferer
whether it is safe to transmit. The third option is via an existing
backhaul which corresponds to methods 4-6 above.

\paragraph*{Enhanced heuristic: Resource block prioritization:} 
Note that throughout this section we have assumed that attempts by a
link to access resource block $c$ are governed by $\lambda_{\ell,r,c}$
and $\mu_{\ell,r,c}$ and are performed independently across resource
blocks. However, since in heterogeneous networks we have two classes
of basestations (macros and femtos) there is potential to reduce
interference if each class has priority on a different set of resource
blocks. We now describe a heuristic to achieve this. In particular if
basestation $i$ is a macro we bias it towards low numbered resource
blocks by only letting one of its users be active on a resource block
if it also has active users on all lower numbered resource
blocks. More formally, if $\ell\in U_i$ then $\langle \ell,r,c\rangle$
is feasible
if for all $c'<c$ there exists $\ell'\in U_i$
and $r'\in \cR$ such that $y_{\ell',r',c'}=1$. Similarly, if
basestation $i$ is a femto we bias it towards high numbered resource
blocks by only letting one of its users be active on a resource block
if it also has active users on all higher numbered resource
blocks. More formally, if $\ell\in U_i$ then $\langle \ell,r,c\rangle$
is feasible if for all $c'>c$ there exists $\ell'\in U_i$
and $r'\in \cR$ such that $y_{\ell',r',c'}=1$. 


\section{Simulation Results}
\label{s:sim}

\begin{figure}
\begin{center}
\includegraphics[width=3.2in]{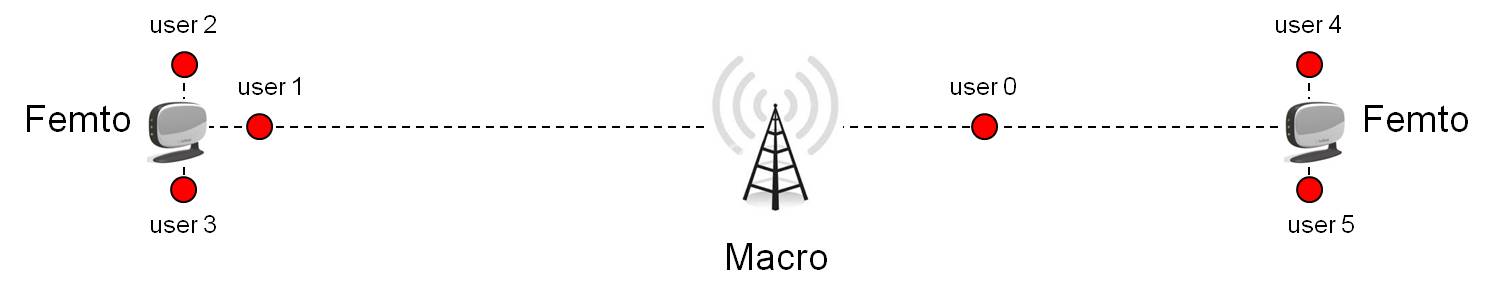}
\caption[]{Network configuration: 1 macro and 2 femtos each with 2 users.}
\label{f:setup}
\end{center}
\end{figure}

We now provide an example to show how the algorithms work.  We
consider a simple toy example since it allows us to compute the
optimal schedule.  We consider three omnidirectional basestations, 1
macro and 2 femtos, together with six users, two for each basestation.
The two macro users are at distance 100m (user 0) and 780m (user 1)
respectively. Each femto has two users (users 2-5) at distance 10m.
The exact configuration together with the user numbering is shown in
Figure~\ref{f:setup}.  We assume that user 1 has to be associated with
the macro since the nearby femto is in CSG mode. The transmit power of
the macro is 46dBm and for the femtos it is 8dBm.  The system
bandwidth is 5MHz and the noise density is -165dBm/Hz. We split the
system bandwidth into three resource blocks. The pathloss is
represented by a COST-231 Hata model. In particular the path loss at
distance $d$ meters is assumed to be $0.525*d^{-3.523}$. For
simplicity we consider two instantaneous transmission rates, a ``low''
rate of 8bits/sec/Hz and a ``high'' rate of 16bits/sec/Hz.  We
consider 
blocks.  

For this configuration we aim to achieve user throughputs that solve
problem (\ref{eq:maxutil2}). We can compute the following
optimal solution offline via a standard subgradient algorithm. For
brevity we use notation of the form $0h2\ell4\ell$ to represent a
schedule, which means  user $0$ is receiving data at the high rate
and users $2$ and $4$ are receiving data at the low rate. 
In the optimal solution for schedules $m\in
\{0h2\ell4\ell,0h3\ell4\ell,0h2\ell5\ell,0h3\ell5\ell\}$ we have
$\pi_m=8.4\%$, for schedules $m\in\{ 1h4\ell,1h5\ell\}$ we have
$\pi_m=10.7\%$, and for schedules $m\in\{2h4h,2h5h,3h4h,3h5h\}$ we
have $\pi_m=11.2\%$.  Figure~\ref{f:ideal} shows the corresponding optimal link throughputs as computed by the subgradient algorithm.
\begin{figure}
\begin{center}
\includegraphics[width=3.4in]{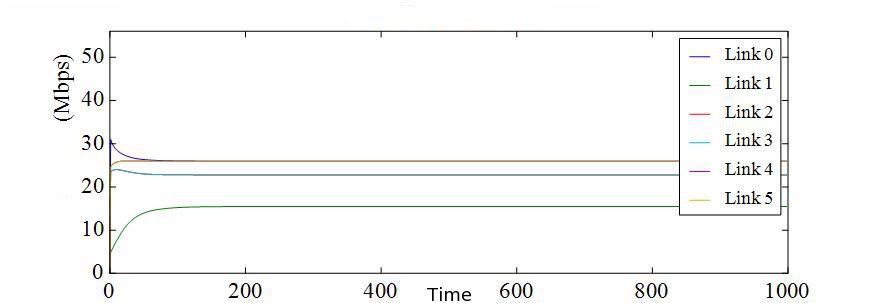}
\caption[]{Link throughputs under the optimal solution. Top curves for users/links 
0, 4 and 5; middle curves for users 2 and 3; bottom curve for user 1. (Note that some curves coincide.)}
\label{f:ideal}
\end{center}
\end{figure}

\begin{figure}
\begin{center}
\includegraphics[width=3.2in]{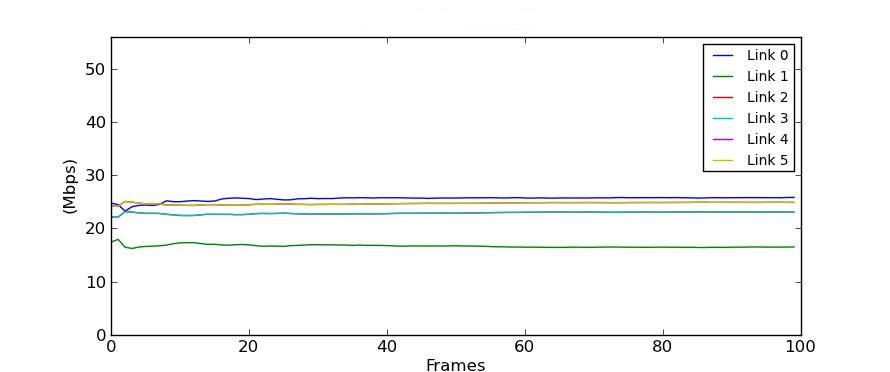}
\caption[]{Link throughputs under MMUO in combination with PF. Top curve for user 
0; second curve for users 4 and 5; third curve for users 2 and 3; bottom curve for user 1.}
\label{f:propfair}
\end{center}
\end{figure}

The behavior of MMUO and MMUO with Proportional Fair are similar.
In the interest of space, we present the plot for MMUO with PF only (in 
Figures~\ref{f:propfair}). 
As we can see, the link throughputs closely approximate the optimal
rates in Figure~\ref{f:ideal}.  In both plots, user 0 (the closer user
to the macro basestation) has the highest throughput while user 1 (the
far user to the macro basestation) has the lowest throughput. Among
the femto users, users 4 and 5 have higher throughputs since they
create less interference to a macro user than users 2 and 3.


We conclude with a brief discussion of how the algorithms for testing
feasibility work in this context. In particular suppose that we are in
configuration $1h4\ell$ and suppose that user $3$ wants to transmit at
the high rate. This is infeasible. User $3$ may discover this by
either a) overhearing the CQI reported by user $1$ and realizing it is
sufficiently close to the minimum acceptable CQI or b) briefly probing
the channel at the high rate and then discovering that user can no
longer support its current rate. In both cases user $3$ decides not to
transmit.
\cm{
\begin{figure}
\begin{center}
\includegraphics[width=3in]{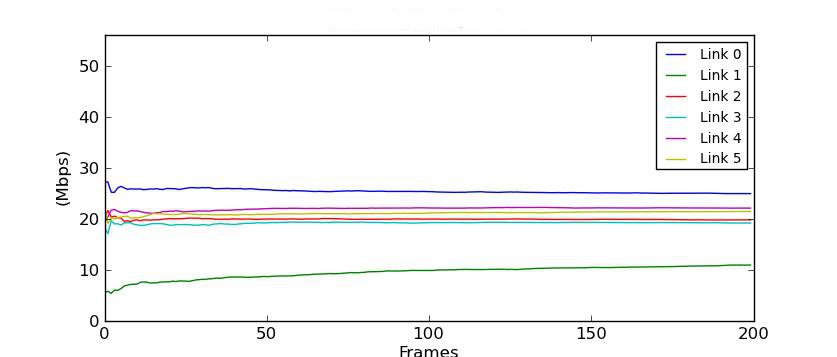}
\caption[]{Link throughputs under the MMUO algorithm. Top curve for users 
0, 4 and 5; middle curve for users 2 and 3; bottom curve for user 1.}
\label{f:MMUO}
\end{center}
\end{figure}
}

\cm{

\begin{figure*}
\begin{center}
\begin{minipage}[t]{3.in}
\includegraphics[width=3in]{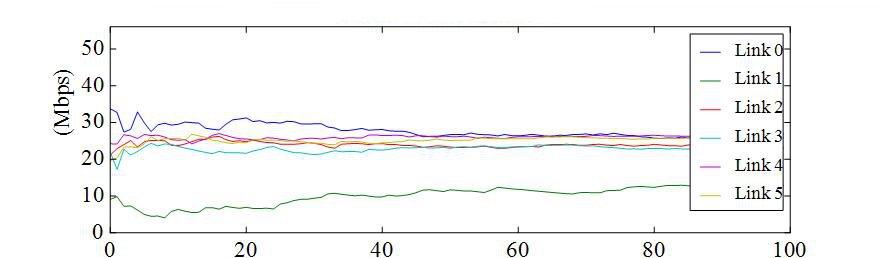}
\end{minipage}
\begin{minipage}[t]{3.in}
\includegraphics[width=3in]{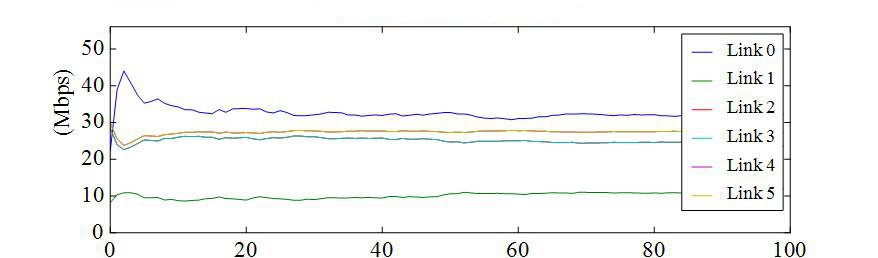}
\end{minipage}
\end{center}
\label{f:ideal}
\end{figure*}

{\bf does it fit here or does it go?}
\paragraph*{Benefits of multiple carriers when applying CSMA} 
One objection that is sometimes raised to applying utility maximal
CSMA to practical systems is that by purely relying on a random-access
type of mechanism, it can take a long time to jump from one maximal
configuration to another and this may be necessary to reach extreme
points of the feasible region.  This in turn may mean that some
links are starved for long periods of time. For example consider a set
of transmitters in a line such that neighboring transmitters
interfere.  In an optimal schedule on a single carrier system we need
to switch from every even-numbered transmitter being active to every
odd-numbered transmitter being active. Using CSMA this switch can take
a long time which will lead to bursty service. However, with multiple
carriers this is much less of an issue since different carriers can be
in different maximal configurations which means that service to
individual links is smoothed out. 
}

\section{Previous Work}
\label{s:otherwork}
We now describe how our work relates to existing techniques. Prior
work mainly falls into two categories, resource allocation in OFDM
systems
and CSMA-based algorithms for 802.11 networks. As we have
seen, our proposal has been to derive an algorithm based on CSMA
techniques for the case of OFDM resource allocation.

\paragraph*{OFDM resource allocation}
LTE uses an OFDM physical layer.  Resource allocation in OFDM systems
addresses problems such as channel selection, local scheduling, power
control and user association, i.e.\ which basestation serves which 
user. One popular technique, e.g.\ used in \cite{ChenBR11,BorstMS11,HouG11}, 
is a {\em Gibbs sampler} approach based on Interacting
Particle Systems. 
The main idea here is that for a given network
configuration each node has a local energy based on the interference
that it both causes and receives. Nodes then pick new states based on
their local energy. Gibbs sampler techniques have also been used to
motivate greedy algorithms for LTE resource block selection,
e.g.~\cite{AndrewsCFG10}. 
Another popular technique, e.g. used in
\cite{StolyarV09,SonLYC12}, is to set power levels according to a gradient
ascent approach. In particular each transmitter adjusts power levels
so as to improve network utility in its neighborhood. Both the Gibbs
sampler and the gradient ascent based methods require information
exchange on how much interference each transmitter causes to each
receiver. For the Gibbs sampler methods interference information needs
to be exchanged in order to calculate local energy levels.  For the
gradient ascent methods nodes need to exchange ``partial derivative''
information to indicate how the interference they experience would be
affected by a change in a neighbor's power levels.  We remark that
MMUO does not require such detailed information exchange.  It bases
its calculations on CQI messages that are already included in LTE, augmented with the activity indicators (and possibly safety margins).

\paragraph*{CSMA-based Algorithms}
In the classic CSMA setup all links wish to access a single channel.
Jiang and Walrand \cite{JiangW08} showed that CSMA can achieve any set
of feasible throughputs.  Since this result, a number of papers have
looked at how to make channel access rates dependent on local queue
sizes in order to keep the system stable,
e.g.~\cite{GhaderiS10,GhaderiS12,BorstGW12,RajagopalanSS09,ShahS10}. As
already discussed, we have based our analysis on the work \cite{LiuYPCP10} (later extended in \cite{ProutiereYLC10,LeeYCNKC12})
that analyzed utility maximization in a CSMA
setting.

\section{Conclusion}
In this paper we have presented a CSMA-based scheduling algorithm for
heterogenous LTE networks with both macro and small cells.  Our
main contribution is twofold.  Mathematically, our algorithm handles
the general multiple transmission rates on multiple carriers and achieves utility
optimality. For the practical setting, the communication among the
basestations utilizes the existing CQI-based technology and hence the
additional signaling is minimal.

\bibliographystyle{plain} \bibliography{/home/ylz/mainrefs}

\end{document}